\documentclass[aps,preprint,nofootinbib,showpacs,superscriptaddress]{revtex4}
\usepackage{amssymb}
\usepackage{amsmath}
\usepackage{graphicx}
\usepackage{multirow}
\usepackage{subfigure}
\usepackage{float}
\usepackage{graphics}
\usepackage{slashed}

\def\be{\begin{equation}}
\def\ee{\end{equation}}
\newcommand{\bea}{\begin{equation} \begin{array}{c}}
\newcommand{\eea}{ \end{array} \end{equation}}

\def\gum{\gamma^{\mu}}

\def\gdm{\gamma_{\mu}}

\def\as{\alpha_s}

\def\nno{\nonumber}

\begin{document}

\title{Model independent analysis of top quark forward-backward asymmetry
 at the Tevatron up to $\mathcal{O}(\as^2/\Lambda^2)$}
\author{Ding Yu Shao}
\affiliation{Department of Physics and State Key Laboratory of
Nuclear Physics and Technology, Peking University, Beijing 100871,
China}
\author{Chong Sheng Li}
\email{csli@pku.edu.cn} \affiliation{Department of Physics and State
Key Laboratory of Nuclear Physics and Technology, Peking University,
Beijing 100871, China} \affiliation{Center for High Energy Physics,
Peking University, Beijing 100871, China}
\author{Jian Wang}
\affiliation{Department of Physics and State Key Laboratory of
Nuclear Physics and Technology, Peking University, Beijing 100871,
China}
\author{Jun Gao}
\affiliation{Department of Physics and State Key Laboratory of
Nuclear Physics and Technology, Peking University, Beijing 100871,
China}
\author{Hao Zhang}
\affiliation{Department of Physics and State Key Laboratory of
Nuclear Physics and Technology, Peking University, Beijing 100871,
China}
\author{Hua Xing Zhu}
\affiliation{Department of Physics and State Key Laboratory of
Nuclear Physics and Technology, Peking University, Beijing 100871,
China}


\begin{abstract}
We present the complete calculations of the forward-backward
asymmetry ($A_{\rm FB}$) and the total cross section of top quark
pair production induced by dimension-six four quark operators at the
Tevatron up to $\mathcal{O}(\as^2/\Lambda^2)$. Our results show that
next-to-leading order (NLO) QCD corrections can change $A_{\rm FB}$
and the total cross section by about 10\%. Moreover, NLO QCD
corrections reduce the dependence of $A_{\rm FB}$ and total cross
section on the renormalization and factorization scales
significantly. We also evaluate the total cross section and the
charge asymmetry ($A_{\rm C}$) induced by these operators at the
Large Hadron Collider (LHC) up to $\mathcal{O}(\as^2/\Lambda^2)$,
for the parameter space allowed by the Tevatron data. We find that
the value of $A_{\rm C}$ induced by these operators is much larger
than SM prediction, and LHC has potential to discover these NP
effects when the measurement precision increases.
\end{abstract}

\pacs{14.65.Ha, 12.38.Bx, 12.60.-i}

\maketitle

\section{INTRODUCTION}\label{s1}
The top quark is the heaviest particle discovered so far, with a
mass close to the electroweak symmetry breaking scale. Thus it is a
wonderful probe for the electroweak breaking mechanism and new
physics (NP) beyond the standard model (SM) through its productions
and decays at colliders. The forward-backward asymmetry ($A_{\rm
FB}$) of the top quark pair production is one of the interesting
observables at hadron colliders. Within the SM, $A_{\rm FB}$ is
absent at the tree level in QCD due to charge symmetry, and occurs
at next-to-leading order (NLO) $\mathcal{O}(\as^3)$ in QCD with the
prediction $A_{\rm FB} \sim 6\%$ in the $t\bar{t}$ rest frame
\cite{Kuhn:1998jr,Kuhn:1998kw,Bowen:2005ap,Almeida:2008ug,Antunano:2007da,Ahrens:2011uf}.
In the last few years, D\O ~and CDF Collaborations measured $ A_{\rm
FB}$ at the Tevatron
\cite{:2007qb,Aaltonen:2008hc,Aaltonen:2011kc,collaboration:2011gf}.
Recently, the CDF Collaborations annouced that, for the invariant
mass of the top quark pair $m_{t\bar t}\geq 450$~GeV, the measured
asymmetry, $A_{\rm FB}=0.475\pm0.114$\cite{Aaltonen:2011kc}, differs
by 3.4$\sigma$ from the SM predictions $A_{\rm FB}=0.088\pm0.013$,
which has aroused many discussions of explaining this deviation in
NP model, including new gauge bosons, axigluons and so
on\cite{Djouadi:2009nb,Jung:2009jz,Cheung:2009ch,Frampton:2009rk,
Shu:2009xf,Arhrib:2009hu,Ferrario:2009ee,Dorsner:2009mq,Jung:2009pi,
Cao:2009uz,Barger:2010mw,Cao:2010zb,Xiao:2010hm,Martynov:2010ed,
Chivukula:2010fk,Bauer:2010iq,Chen:2010hm,Jung:2010yn,Burdman:2010gr,
Jung:2010ri,Choudhury:2010cd,Cheung:2011qa,Cao:2011ew,Berger:2011ua,
Barger:2011ih,Bhattacherjee:2011nr,Blum:2011up,Patel:2011eh,
Isidori:2011dp,Zerwekh:2011wf,Barreto:2011au,Foot:2011xu,Ligeti:2011vt,
Gresham:2011pa,Jung:2011zv,Buckley:2011vc,Shu:2011au,AguilarSaavedra:2011zy,
Chen:2011mga,Degrande:2011rt,Jung:2011ua,Jung:2011ue,
Babu:2011yw,Djouadi:2011aj,Barcelo:2011fw,Krohn:2011tw,AguilarSaavedra:2011hz,
Cui:2011xy,Hektor:2011ms,Gabrielli:2011jf,Duraisamy:2011pt,
AguilarSaavedra:2011ug,Barcelo:2011vk,Tavares:2011zg,Vecchi:2011ab,Blum:2011fa,
Degrande:2010kt,AguilarSaavedra:2011vw}. \\
\indent Since we do not know which type of NP will be responsible
for this deviation, it is interesting to study the $A_{\rm FB}$ in a
model independent way, using an effective Lagrangian. In general, NP
scale relevant to $A_{\rm FB}$ is large enough so that the heavy
fields have been integrated out at the low energy scale. At the
Tevatron, the subprocess $q\bar q\to t\bar t$ dominates over top
quark pair production, so only contributions from dimension-six four
quark operators to the $t\bar t$ production are considered. Similar
approach had been adopted for the dijet production to constrain the
composite scale of light quarks~\cite{Eichten:1983hw,Eichten:1984eu,
Chiappetta:1990jd,Eichten:1996xn,Gao:2011ha}. The relevant effective
Lagrangian can be written as
\begin{eqnarray}\label{lnp}
  \mathcal {L}_{NP}=\frac{1}{\Lambda^2}\sum_{A,B}\left[C^{1}_{AB}\left(\bar{q}_A\gdm q_A\right)\left(\bar{t}_B\gum t_B\right)+
   C^{8}_{AB}\left(\bar{q}_A T^a\gdm q_A\right)\left(\bar{t}_B T^a\gum
   t_B\right)\right],
\end{eqnarray}
where $\{A,B\}=\{L,R\}$ with $q=(u,d)^T, (c,s)^T$. Up to
$\mathcal{O}(\as/\Lambda^2)$, the NP contributions to the total
cross section and the $A_{\rm FB}$ are clear in the vector-axial
basis~\cite{Degrande:2010kt,Blum:2011up,Kamenik:2011wt}, as compared
with the chirality basis. Only the axial-axial current combination
contributions to the $A_{\rm FB}$, and the vector-vector operator
contributes to the total cross section. However, this is no longer
true up to $\mathcal{O}(\as^2/\Lambda^2)$. The chirality basis is
the preferred one when studying the chiral structure of NP effects
much above the Electroweak scale, so we choose to work in the
chirality basis. The contributions to $A_{\rm FB}$ at leading order
(LO) from such operators has been explored in
Refs.~\cite{Jung:2009pi,Jung:2010yn,Jung:2010zzd,Jung:2010ri,
Degrande:2010kt,AguilarSaavedra:2011vw}. It is shown that the
$A_{\rm FB}$ observed at the Tevatron can be explained by above
operators for suitable parameters. As we know, the LO cross section
at hardron colliders suffers from large uncertainties due to the
arbitrary choice of the renormalization scale and factorization
scale, thus it is important to include NLO corrections to improve
theoretical predictions. Besides, at the NLO level, virtual
corrections, real gluon emission and massless (anti)quark emission
can lead to a sizeable difference between the differential top and
anti-top production process~\cite{Kuhn:1998jr,Kuhn:1998kw}, which
will contribute to
$A_{\rm FB}$.\\
\indent In this paper, we present the complete NLO QCD calculations
of $A_{\rm FB}$ and the total cross section of top quark pair
production at the Tevatron induced by above operators, and we also
study the top quark pair production at the Large Hadron Collider
(LHC) induced by these operators at the NLO QCD level. Last year,
LHC reported their first observation of top quark pair production,
and will soon become a major top quark factory. At the LHC, the top
quark pairs can be produced through quark antiquark annihilation
$q\bar q\to t\bar t$ and gluon fusion $gg\to t\bar t$. Since gluon
fusion channel dominates at the LHC, it is difficult to probe these
four quark effective operators from early LHC results. However, it
is still possible to detect these effects from above effective
operators on the Charge Asymmetry($A_{\rm C}$) at the LHC, in the
parameter space allowed by the Tevatron data, when the measurement
precision increases.\\
\indent The arrangement of this paper is as follows. In
Sec.~\ref{s2} we show the LO results. In Sec.~\ref{s3}, we present
the details of the NLO calculations, including the virtual and real
corrections to the top quark pair production. Section~\ref{s4}
contains the numerical results, and Section~\ref{s5} is a brief
summary.
\section{LO results}\label{s2}
Throughout our calculation, we adopt the same conventions as in
Ref.~\cite{Zhu:2011gd} (see Sec. III A), and present the helicity
amplitudes for $q\bar q\to t\bar t$ in the Four-Dimensional Helicity
(FDH) regularization scheme~\cite{Bern:2002zk}. The $t\bar t$
production amplitudes, including NP contributions, can be written as
\begin{eqnarray}
  \mathcal{M}_{t\bar t}=\as f_{\rm LO}^{\rm SM}+\frac{1}{\Lambda^2}f_{\rm LO}^{\rm NP} +
  \as^2 f_{\rm NLO}^{\rm SM} + \frac{\as}{\Lambda^2}f_{\rm NLO}^{\rm
  NP}+\cdots,
\end{eqnarray}
and thus the partonic cross section, up to
$\mathcal{O}(\as^2/\Lambda^2)$, can be written as
\begin{eqnarray}
  \hat{\sigma}_{t\bar t}&=&\as^2f_{\rm LO}^{\rm SM}f^{\rm SM^{\ast}}_{\rm LO} +
  2\frac{\as}{\Lambda^2}\mathcal{R}e
  \left(f_{\rm LO}^{\rm SM}f^{\rm NP^{\ast}}_{\rm LO}\right) \nno\\
  &&+2\as^3
  \mathcal{R}e\left( f_{\rm LO}^{\rm SM}f^{\rm SM^{\ast}}_{\rm NLO} \right) + 2
  \frac{\as^2}{\Lambda^2} \left[ \mathcal{R}e\left( f_{\rm LO}^{\rm NP}f^{\rm SM^{\ast}}_{\rm NLO}
  \right) + \mathcal{R}e\left( f_{\rm LO}^{\rm SM}f^{\rm NP^{\ast}}_{\rm NLO}
  \right)\right].
\end{eqnarray}
\begin{figure}[h!]
\begin{centering}
\includegraphics[scale=0.8]{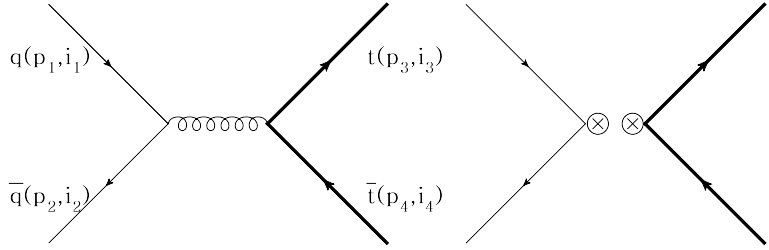}
\par\end{centering}
\caption{\label{fig:LO feynman diagrams}LO Feynman diagrams for
$q\bar{q}\rightarrow t\bar{t}$ induced by SM QCD and NP
interactions.}
\end{figure}
\indent  The LO Feynman diagrams for the subprocess $q(p_1)\bar
q(p_2)\rightarrow t(p_3)\bar t(p_4)$ induced by the SM QCD and the
NP interactions are shown in Fig.~\ref{fig:LO feynman diagrams}, and
their ($+-++$) helicity amplitudes are
\begin{eqnarray}
  \mathcal{M}_{\textrm{LO}}^{\textrm{SM}}(+-++)&=&\frac{i8\pi\as m_t}{s}\left(\mathcal{M}_1+\mathcal{M}_2\right)\mathcal{C}_8,\\
  \mathcal{M}_{\textrm{LO}}^{\textrm{NP}}(+-++)&=&\frac{i2m_t}{\Lambda^2}\left[\left(\mathcal {M}_1 C_{\rm RR}^1 +
  \mathcal {M}_2 C_{\rm RL}^1 \right)\mathcal{C}_1 + \left(\mathcal {M}_1 C_{\rm RR}^8 + \mathcal
  {M}_2 C_{\rm RL}^8\right)\mathcal{C}_8\right],
\end{eqnarray}
where the SM QCD and NP contributions are denoted by superscipts SM
and NP, and Mandelstam variables $s$, $t$ and $u$ are defined as
follows:
\begin{eqnarray}
  s=(p_1 + p_2)^2,~~~t=(p_1 - p_3)^2,~~~u = (p_1 - p_4)^2.
\end{eqnarray}
We define the following abbreviations for the color structures and
matrix elements,
\begin{eqnarray}
   \mathcal{M}_1=\frac{\langle\eta_4 1\rangle\langle \eta_3|\textbf{3}|2]}{\langle 3^{\flat}\eta_3\rangle\langle \eta_4
  4^{\flat}\rangle},&~~~~&
  \mathcal{M}_2=\frac{\langle\eta_3 1\rangle\langle \eta_4|\textbf{4}|2]}{\langle 3^{\flat}\eta_3\rangle\langle \eta_4
  4^{\flat}\rangle},\nno\\
  \mathcal{M}_3=\frac{\langle\eta_4 2\rangle\langle \eta_3|\textbf{3}|1]}{\langle 3^{\flat}\eta_3\rangle\langle \eta_4
  4^{\flat}\rangle},&~~~~&
  \mathcal{M}_4=\frac{\langle\eta_3 2\rangle\langle \eta_4|\textbf{4}|1]}{\langle 3^{\flat}\eta_3\rangle\langle \eta_4
  4^{\flat}\rangle},\nno\\
  \mathcal{C}_1=\delta_{i_2 i_1}\delta_{i_3 i_4},&~~~~&
  \mathcal{C}_8=T^a_{i_2 i_1}T^a_{i_3 i_4},
\end{eqnarray}
where $i_{1...4}$ are the color indices of the external quarks and
the boldface momenta denotes massive vectors. We use the modified
spinor helicity method suited for massive particles
\cite{Kleiss:1985yh} in our calculations, and a recent application
of this method can be found in the Ref~\cite{Badger:2010mg}. The
($-+++$) amplitudes are given by
\begin{eqnarray}
  \mathcal{M}_{\rm{LO}}^{\rm{SM}}(-+++)&=&\frac{i8\pi\as m_t}{s}\left(\mathcal{M}_3+\mathcal{M}_4\right)\mathcal{C}_8,\\
  \mathcal{M}_{\rm{LO}}^{\rm{NP}}(-+++)&=&\frac{i2m_t}{\Lambda^2}\left[\left(\mathcal {M}_3 C_{\rm LR}^1 +
  \mathcal {M}_4 C_{\rm LL}^1 \right)\mathcal{C}_1 + \left(\mathcal {M}_3 C_{\rm LR}^8 + \mathcal
  {M}_4 C_{\rm LL}^8\right)\mathcal{C}_8\right].
\end{eqnarray}
The amplitudes with other helicity configurations can be obtained
from ($+-++$) and ($-+++$) by exchanging light-like momenta
$p^{\flat}$ and $\eta$ \cite{Zhu:2011gd,Badger:2010mg}. At the LO,
there is only vector current coupling $\bar{\psi}\gamma^{\mu}\psi$
at the massive quark vertex. At the NLO, however, magnetic-momentum
coupling $\bar{\psi}(i\sigma^{\mu\nu }(p_3 + p_4)_{\mu})\psi/(2m_t)$
is induced from loop diagrams. For completeness we list the matrix
elements for magnetic-moment interaction,
\begin{eqnarray}
\mathcal{M}_1^{(m)}=\frac{m_t^2\langle\eta_3 1\rangle\langle \eta_4
1\rangle[21]}{\langle 3^{\flat}\eta_3\rangle\langle \eta_4
  4^{\flat}\rangle},&~~~~&
  \mathcal{M}_2^{(m)}=\frac{\langle1 2\rangle\langle \eta_3|\textbf{3}|2]\langle \eta_4|\textbf{4}|2]}
  {\langle 3^{\flat}\eta_3\rangle\langle \eta_4 4^{\flat}\rangle},\nno\\
  \mathcal{M}_3^{(m)}=\frac{m_t^2\langle\eta_3 2\rangle\langle \eta_4 2\rangle[21]}{\langle 3^{\flat}\eta_3\rangle\langle \eta_4
  4^{\flat}\rangle},&~~~~&
  \mathcal{M}_4^{(m)}=\frac{\langle1 2\rangle\langle \eta_3|\textbf{3}|1]\langle \eta_4|\textbf{4}|1]}
  {\langle 3^{\flat}\eta_3\rangle\langle \eta_4
  4^{\flat}\rangle}.
\end{eqnarray}

After phase space integration, the $\mathcal{O}(\as/\Lambda^2)$
partonic differential cross section is
\begin{eqnarray}\label{treehard}
  \frac{d\hat{\sigma}^{\textrm{NP}}_{\rm{LO}}}{d
  cos\theta}&=&\frac{\beta}{18}\frac{\as}{\Lambda^2}\left[ \frac{1}{4}(1+ \rho + \beta^2cos^2\theta)
  (C^{8}_{\rm LR}+C^{8}_{\rm RR})+\frac{1}{2}\beta\cos\theta(C^8_{\rm RR}-C^8_{\rm LR})
  \right],
\end{eqnarray}
where $\rho=4m_t^2/s$, $\beta=\sqrt{1-\rho^2}$, and $\theta$ is the
polar angle between the incoming quark and the outgoing top quark in
the $t\bar t$ rest frame. The color and spin indices are
averaged(summed) over initial(final) states. In Eq.~(\ref{treehard})
the term linear in $\cos\theta$ could generate $A_{\rm FB}$
proportional to $\rm \left(C^8_{RR}-C^8_{LR}\right)$ and the rest
terms contribute to the total cross section proportional to $\rm
\left(C^8_{RR} +
C^8_{LR}\right)$. These relations will be changed at the NLO level.\\
\indent The LO total cross section at the hadron collider is
obtained by convoluting the partonic level cross section with the
Parton Distribution Function (PDF) $f_{i/A}$ for the initial hadron
A:
\begin{eqnarray}
  \sigma_{\rm{LO}}=\sum_{a,b}\int_{\tau}^1dx_a\int_{\tau/x_a}^1dx_b
  f_{a/A}(x_a,\mu_f)f_{b/B}(x_b,\mu_f)\hat{\sigma}_{\rm{LO}},
\end{eqnarray}
where $\tau=4m_t^2/s$. The sum is over all possible initial partons.

\section{NLO QCD corrections}\label{s3}
The NLO corrections to the top pair production consist of the
virtual corrections, generated by loop diagrams of colored
particles, and the real corrections with the radiation of a real
gluon or a massless (anti)quark. We carried out all the calculations
in the 't Hooft-Feynman gauge and used the FDH scheme to regularize
all the divergences. Moveover, for the real corrections, we used the
dipole substraction method with massive partons~\cite{Catani:2002hc}
to separate the infrared (IR) divergences, which is convenient for
the case of massive Feynman diagrams and provides better numerical
accuracy.

\subsection{Virtual corrections}
\begin{figure}[h!]
\begin{centering}
\includegraphics[scale=1]{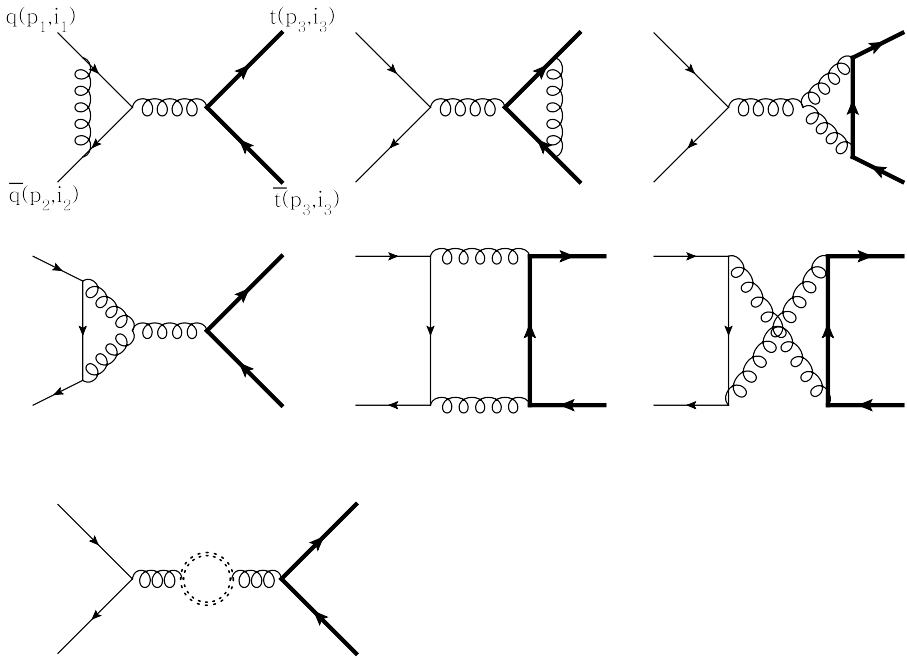}
\par\end{centering}
\caption{\label{fig:smloop}One-loop virtual Feynman diagrams for
$q\bar{q}\rightarrow t\bar{t}$ induced by SM QCD interactions. }
\end{figure}

\begin{figure}[h!]
\begin{centering}
\includegraphics[scale=1]{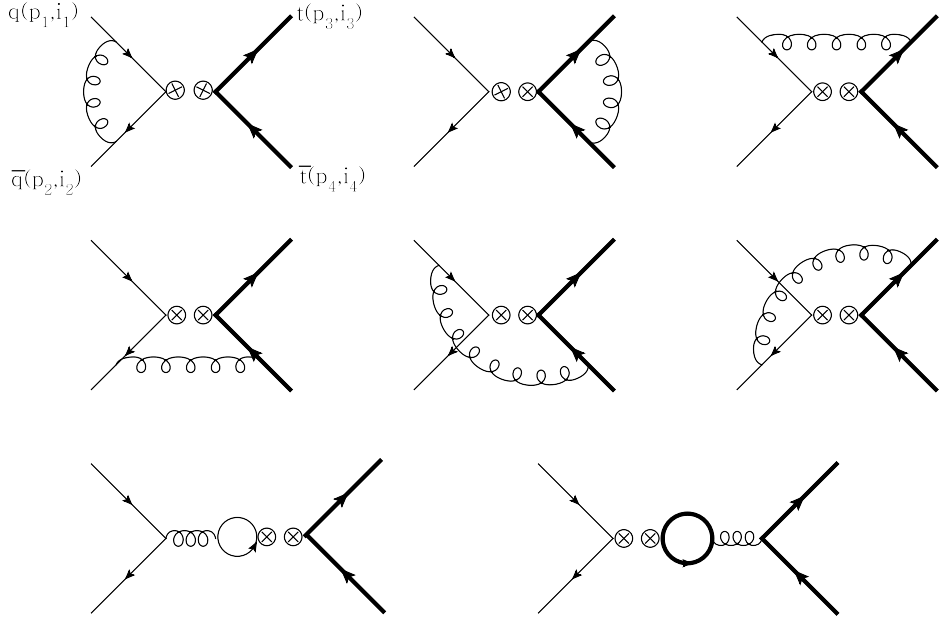}
\par\end{centering}
\caption{\label{fig:NPvirtual}One-loop virtual Feynman diagrams for
$q\bar{q}\rightarrow t\bar{t}$ induced by NP interactions. }
\end{figure}

The virtual corrections for the top quark pair production include
the box diagrams, triangle diagrams, and self-energy diagrams in SM
QCD and NP as shown in Fig.~\ref{fig:smloop} and
Fig.~\ref{fig:NPvirtual}. We have calculated the one-loop helicity
amplitudes for the SM process, and find complete agreement with
those in the Ref.~\cite{Zhu:2011gd,Korner:2002hy}. Here we only list
the NP contributions.

All the ultraviolet (UV) divergences in the loop diagrams are
canceled by counterterms for the wave functions of the external
fields ($\delta Z_q,\delta Z_t$), and the Wilsion coefficients
$\delta Z_{C^i_{AB}}$. For the external fields, we fix all the
renormalization constants using on-shell subtraction, and,
therefore, they also have IR singularities
\begin{eqnarray}
  \delta Z_q &=& -\frac{\as}{4\pi}C_{\varepsilon}C_{F}
  \left(\frac{1}{\varepsilon_{UV}}-\frac{1}{\varepsilon_{IR}}\right),\\
  \delta Z_t &=& -\frac{\as}{4\pi}C_{\varepsilon}C_{F}
  \left(\frac{1}{\varepsilon_{UV}}+\frac{2}{\varepsilon_{IR}}+5-3\ln\frac{m_t^2}{\mu_r^2}\right),
\end{eqnarray}
where
$C_{\varepsilon}=(4\pi)^{\varepsilon}\frac{1}{\Gamma(1-\varepsilon)}$
, $C_F=4/3$ and $\mu_r$ is the renormalization scale. For
contourterms of the Wilsion coefficients $\delta Z_{C^i_{AB}}$, we
adopted the $\rm \overline{MS}$ scheme
\begin{eqnarray}
\delta Z_{C^i_{AB}}=\frac{\as}{4\pi}C_{\varepsilon} C_F
\frac{1}{\varepsilon_{UV}}\left(
\begin{array}{cccccccc}
 0 & \frac{9}{8} & 0 & 0 & 0 & 0 & 0 & 0 \\
 \frac{1}{4} & \frac{n_f-5}{16} & 0 & \frac{1}{16} & 0 & \frac{n_f}{16} & 0 & 0 \\
 0 & 0 & 0 & -\frac{9}{8} & 0 & 0 & 0 & 0 \\
 0 & \frac{1}{16} & -\frac{1}{4} & \frac{n_f-20}{16} & 0 & 0 & 0 & \frac{n_f}{16} \\
 0 & 0 & 0 & 0 & 0 & -\frac{9}{8} & 0 & 0 \\
 0 & \frac{n_f}{16} & 0 & 0 & -\frac{1}{4} & \frac{n_f-20}{16} & 0 & \frac{1}{16} \\
 0 & 0 & 0 & 0 & 0 & 0 & 0 & \frac{9}{8} \\
 0 & 0 & 0 & \frac{n_f}{16} & 0 & \frac{1}{16} & \frac{1}{4} & \frac{n_f-5}{16}
\end{array}
\right),
\end{eqnarray}
where $n_f=5$ is the number of massless quarks appearing in the
closed loop diagram, and the order of the Wilsion coefficients is
\begin{eqnarray}
\left(C_{\rm LL}^1,C_{\rm LL}^8,C_{\rm LR}^1,C_{\rm LR}^8,C_{\rm
RL}^1,C_{\rm RL}^8,C_{\rm RR}^1,C_{\rm RR}^8\right).
\end{eqnarray}
We have considered mixing effects of different color and chiral
operators, and the evolution equations of the Wilson coefficients
are given in the Appendix \ref{a2}. The renormalized virtual
amplitudes can be written as
\begin{eqnarray}
  \mathcal{M}^V=\mathcal{M}^{\textrm{unren}}+\mathcal{M}^{\textrm{con}}.
\end{eqnarray}
Here $\mathcal{M}^{\rm unren}$ contains the self-energy and vertex
corrections, and $\mathcal{M}^{\textrm{con}}$ are the corresponding
counterterms. The renormalized amplitude $\mathcal{M}^V$ is UV
finite, but still contains IR divergences, which are given by
\begin{eqnarray}
  \mathcal{M}_{\rm{SM}}^{\rm{IR}}&=&\frac{\as}{2\pi}C_{\varepsilon}
  C_F\left[ f_{\text{os}}^{\rm{IR}} \mathcal{C}_1 \mathcal{M}_8^{\rm{SM}}
  + f_{\text{oo}}^{\rm{IR}} \mathcal{C}_8 \mathcal{M}_8^{\rm{SM}}
  \right],\\
  \mathcal{M}_{\rm{NP}}^{\rm{IR}}&=&\frac{\as}{2\pi}C_{\varepsilon}
  C_F\left[ f_{\text{ss}}^{\rm{IR}}\mathcal{C}_1 \mathcal{M}_1^{\rm{NP}}
  + f_{\text{so}}^{\rm{IR}} \mathcal{C}_8 \mathcal{M}_1^{\rm{NP}} +
  f_{\text{os}}^{\rm{IR}}\mathcal{C}_1\mathcal{M}_8^{\rm{NP}}
  + f_{\text{oo}}^{\rm{IR}}\mathcal{C}_8\mathcal{M}_8^{\rm{NP}}
  \right],
\end{eqnarray}
where we define the IR divergence coefficients
$f_{\text{ss}}^{\text{IR}}$, $f_{\text{so}}^{\text{IR}}$,
$f_{\text{os}}^{\text{IR}}$ and $f_{\text{oo}}^{\text{IR}}$ for
different color configurations, "s" for singlet and "o" for octect,
\begin{eqnarray}
f_{\text{ss}}^{\text{IR}}&=&-\frac{1}{\varepsilon_{\rm{IR}}^2}+
\frac{1}{\varepsilon_{\rm{IR}}}\left[\frac{1}{2}\left(\beta
+\frac{1}{\beta }\right)\ln \left(\frac{\beta +1}{\beta
-1}\right)+\ln
\left(\frac{-s}{\mu_r^2}\right)-\frac{5}{2}\right],\\
f_{\text{so}}^{\text{IR}}&=&\frac{3}{2}\ln
\left(\frac{t_1}{u_1}\right)\frac{1}{\varepsilon_{\rm{IR}}},\\
f_{\text{os}}^{\text{IR}}&=&\frac{1}{3}\ln
\left(\frac{t_1}{u_1}\right)\frac{1}{\varepsilon_{\rm{IR}}},\\
f_{\text{oo}}^{\text{IR}}&=&-\frac{1}{\varepsilon_{\rm{IR}}^2}+
\frac{1}{\varepsilon_{\rm{IR}}}\Bigg[-\frac{1}{16}\left(\beta
+\frac{1}{\beta }\right)\ln \left(\frac{\beta +1}{\beta
-1}\right)-\frac{9}{8}\ln \left(\frac{m_t^2}{\mu_r
^2}\right)-\frac{1}{8}\ln \left(\frac{-s}{\mu_r
^2}\right)\nno\\
&&+\frac{7}{4}\ln \left(\frac{t_1}{\mu_r^2}\right)+\frac{1}{2}\ln
\left(\frac{u_1}{\mu_r^2}\right)-\frac{5}{2}\Bigg],
\end{eqnarray}
where $t_1 = m_t^2 - t$, $u_1 = m_t^2 - u$ and
$\mathcal{M}_1^{\rm{NP}}$, $\mathcal{M}_8^{\rm{NP}}$ and
$\mathcal{M}_8^{\rm{SM}}$ are defined as follows,
\begin{eqnarray}
  \mathcal{M}^{\rm{SM}}_{\rm LO} &=& \mathcal{M}_8^{\rm{SM}}\mathcal{C}_8,\\
  \mathcal{M}^{\rm{NP}}_{\rm LO} &=& \mathcal{M}_1^{\rm{NP}} \mathcal{C}_1 +
  \mathcal{M}_8^{\rm{NP}} \mathcal{C}_8.
\end{eqnarray}
Since we only consider high order corrections up to
$\mathcal{O}(\as^2/\Lambda^2)$, the IR divergences of the virtual
corrections can be written as
\begin{eqnarray}\label{vir}
  &&2\mathcal{R}e\left[\mathcal{M}_{\rm{SM}}^{\rm{IR}}\mathcal{M}_{\rm{NP}}^{\rm{LO}^\ast}\right] +
  2\mathcal{R}e\left[ \mathcal{M}_{\rm{NP}}^{\rm{IR}}\mathcal{M}_{\rm{SM}}^{\rm{LO}^\ast}
  \right]\nno\\
  &=&\frac{\as}{\pi}C_{\varepsilon}
  C_F\left[ \left(9f_{\text{os}}^{\rm{IR}} + 2f_{\text{so}}^{\rm{IR}} \right)\mathcal{R}e\left(
  \mathcal{M}_1^{\rm{NP}^\ast}\mathcal{M}_8^{\rm{SM}}\right) + 4 f_{\text{oo}}^{\rm{IR}}\mathcal{R}e\left(
  \mathcal{M}_8^{\rm{NP}^\ast}\mathcal{M}_8^{\rm{SM}}  \right) \right].
\end{eqnarray}
The finite terms in $\mathcal{M}^V_{\rm NP}$ are given in the
Appendix \ref{a1}.

\subsection{Real corrections}\label{ss1}
\begin{figure}[h!]
\begin{centering}
\includegraphics[scale=1]{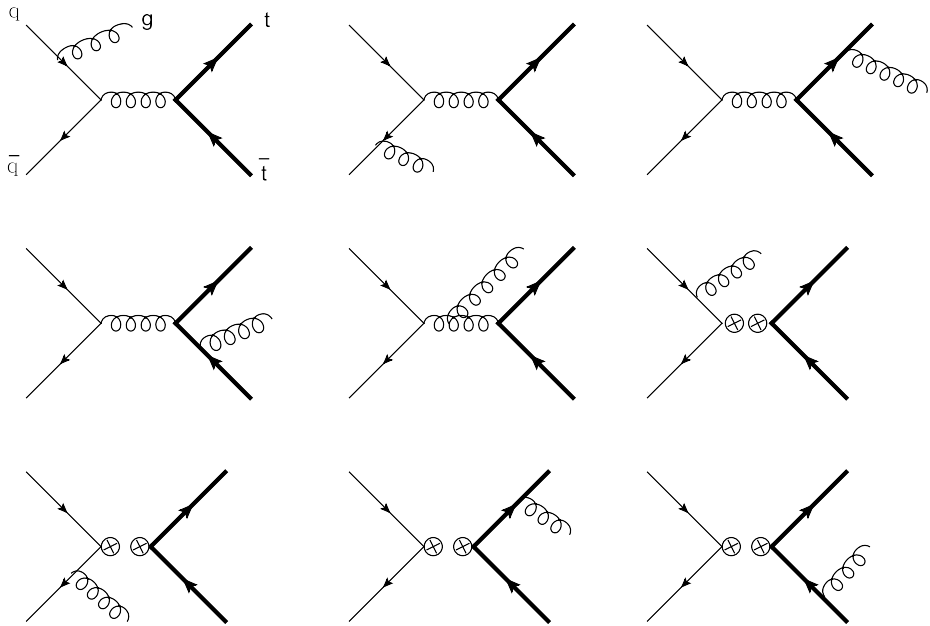}
\par\end{centering}
\caption{\label{fig:rge}Feynman diagrams for the real gluon emission
contributions induced by SM QCD and NP interactions. }
\end{figure}

\begin{figure}[h!]
\begin{centering}
\includegraphics[scale=1]{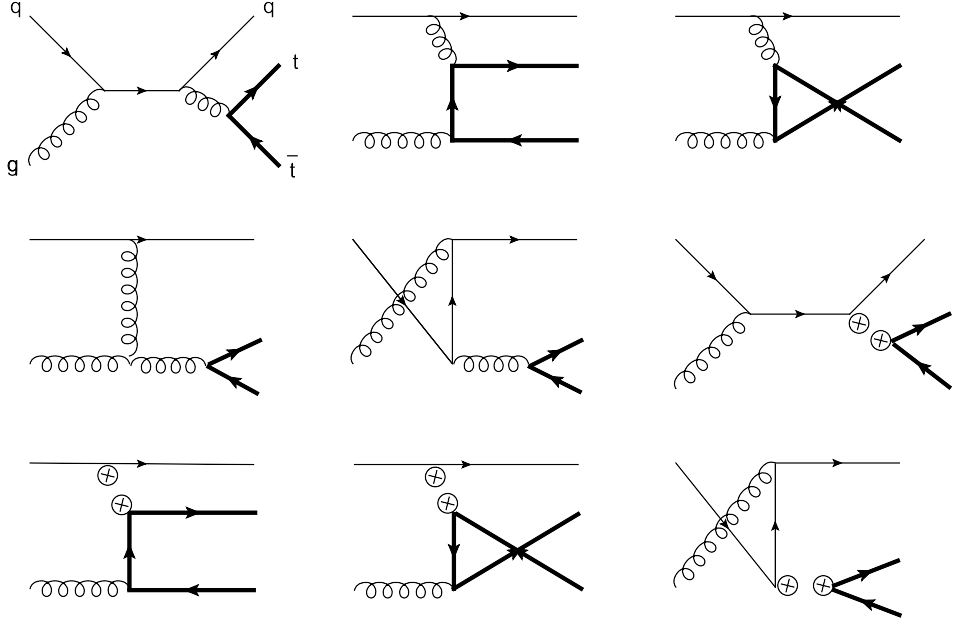}
\par\end{centering}
\caption{\label{fig:rqe}Feynman diagrams for the massless quark
emission contributions induced by SM QCD and NP interactions. }
\end{figure}

At the NLO level the real corrections consist of the radiations of
an additional gluon or massless (anti)quark in the final states,
including the subprocess
\begin{eqnarray}
  q\bar q\to t\bar t g,~~~g q(\bar q)\to t \bar t q(\bar q)
\end{eqnarray}
as shown in Fig.\ref{fig:rge} and Fig.\ref{fig:rqe}.

\indent Before performing the numerical calculations, we need to
extract the IR divergences in the real corrections. In the dipole
formalism this is done by subtracting some dipole terms from the
real corrections to cancel the singularities and large logarithms
exactly, and then the real corrections become integrable in four
dimensions. On the other hand, these dipole subtraction terms are
analytically integrable in $n$ dimensions over one-parton subspaces,
which give $\varepsilon$ poles that represent the soft and collinear
divergences. Then we can add them to the virtual corrections to
cancel the $\varepsilon$ poles, and ensure the virtual corrections
are also integrable in four dimensions. This whole procedure can be
illustrated by the formula~\cite{Catani:2002hc}:
\begin{equation}
\hat{\sigma}^{\rm NLO}=\int_{m+1}\left[\left(d\hat{\sigma}^{\rm{R}}
\right)_{\varepsilon=0}-\left(d\hat{\sigma}^{\rm{A}}\right)_{\varepsilon=0}\right]
+\int_m\left[d\hat{\sigma}^{\rm{V}}+\int_1d\hat{\sigma}^{\rm{A}}\right]_{\varepsilon=0},
\end{equation}
where $m$ is the number of final state particles at the LO, and
$d\hat{\sigma}^{\rm{A}}$ is a sum of the dipole terms. Besides, at
hadron colliders, we have to include the well-known collinear
subtraction counterterms in order to cancel the collinear
divergences arising from the splitting processes of the initial
state massless partons. Here we use the $\overline{\rm MS}$ scheme
and the corresponding NLO PDFs.

For the process with two initial state hadrons, the dipole terms can
be classified into four groups, the final-state emitter and
final-state spectator type,
\begin{eqnarray}
&&\mathcal {D}_{ij,k}(p_1,...,p_{m+1})=\nonumber\\
&&\qquad -{1\over (p_i+p_j)^2-m_{ij}^2}{\
_{m}}\langle...,\widetilde{ij},...,\widetilde
k,...|\frac{\textbf{T}_k\cdot\textbf{T}_{ij}}{\textbf{T}_{ij}^2}
{\textbf{V}_{ij,k}}|..., \widetilde{ij},...,\widetilde
k,...\rangle_{m},
\end{eqnarray}
the final-state emitter and initial-state spectator type,
\begin{eqnarray}
&&\mathcal {D}_{ij}^a(p_1,...,p_{m+1};p_a,...)=\nonumber\\
&&\qquad -{1\over (p_i+p_j)^2-m_{ij}^2}\frac{1}{x_{ij,a}} {\
_{m,a}}\langle...,\widetilde{ij},...;\widetilde
a,...|\frac{\textbf{T}_a\cdot\textbf{T}_{ij}}{\textbf{T}_{ij}^2}
{\textbf{V}_{ij}^a}|..., \widetilde{ij},...;\widetilde
a,...\rangle_{m,a},
\end{eqnarray}
the initial-state emitter and final-state spectator type,
\begin{eqnarray}
&&\mathcal {D}_{j}^{ai}(p_1,...,p_{m+1};p_a,...)=\nonumber\\
&&\qquad -{1\over 2p_a p_i}\frac{1}{x_{ij,a}} {\
_{m,\widetilde{ai}}}\langle...,\widetilde{j},...;\widetilde
{ai},...|\frac{\textbf{T}_j\cdot\textbf{T}_{ai}}{\textbf{T}_{ai}^2}
{\textbf{V}_{j}^{ai}}|..., \widetilde{j},...;\widetilde
{ai},...\rangle_{m,\widetilde{ai}},
\end{eqnarray}
and the initial-state emitter and initial-state spectator type,
\begin{eqnarray}
&&\mathcal {D}^{ai,b}(p_1,...,p_{m+1};p_a,p_b)=\nonumber\\
&&\qquad -{1\over 2p_a p_i}\frac{1}{x_{i,ab}} {\
_{m,\widetilde{ai}}}\langle...;\widetilde
{ai},b|\frac{\textbf{T}_b\cdot\textbf{T}_{ai}}{\textbf{T}_{ai}^2}
{\textbf{V}^{ai,b}}|...;\widetilde {ai},b\rangle_{m,\widetilde{ai}},
\end{eqnarray}
where $a,b$ and $i,j,...$ are the initial and final state partons,
and $\textbf{T}$ and $\textbf{V}$ are the color charge operators and
dipole functions acting on the LO amplitudes, respectively. The
explicit expressions for $x_{i,ab}$, $x_{ij,a}$ and $\textbf{V}$ can
be found in Ref.~\cite{Catani:2002hc}. The integrated dipole
functions together with the collinear counterterms can be written in
the following factorized form
\begin{eqnarray}\label{eq4}
\sim &&\int d\Phi^{(m)}(p_a,p_b) \
_{m,ab}\langle...;p_a,p_b|\textbf{I}_{m+a+b}(\varepsilon)|
...;p_a,p_b\rangle_{m,ab}\nonumber\\
&&+\sum_{a'}\int_0^1 dx\int d\Phi^{(m)}(xp_a,p_b) _{m,a'b}\langle
...;xp_a,p_b|\textbf{P}_{m+b}^{a,a'}
(x)+\textbf{K}_{m+b}^{a,a'}(x)|...;xp_a,p_b\rangle_{m,a'b}
\nonumber\\
&&\qquad \qquad \qquad +(a\leftrightarrow b),
\end{eqnarray}
where $x$ is the momentum fraction of the splitting parton,
$d\Phi^{(m)}$ contains all the factors apart from the squared
amplitudes, $\textbf{I}$, $\textbf{P}$, and $\textbf{K}$ are
insertion operators defined in~\cite{Catani:2002hc}. For simplicity,
in all the above formulas we do not show the jet functions that
define the observables and are included in our numerical
calculations.

The operators $\textbf{P}$ and $\textbf{K}$ provide finite
contributions to the NLO corrections, and only the operator
$\textbf{I}$ contains the IR divergences
\begin{eqnarray}\label{eq3}
\textbf{I}|_{\rm{IR}}&=&-\frac{\alpha_s}{2\pi}{(4\pi)^{\varepsilon}\over
\Gamma(1-\varepsilon)}\bigg\{\sum_j\sum_{k\neq j}\textbf{T}_j\cdot
\textbf{T}_k\bigg[\left(\frac{\mu_r^2}{s_{jk}}\right)^{\varepsilon}
\mathcal{V}(s_{jk},m_j,m_k;\varepsilon_{IR})
+{1\over \textbf{T}_j^2}\Gamma_j(m_j,\varepsilon_{IR})\bigg]\nonumber\\
&& +\sum_j\textbf{T}_j\cdot\textbf{T}_a\bigg[2\left(
\frac{\mu_r^2}{s_{ja}}\right)^{\varepsilon}\mathcal{V}(s_{ja},m_j,0;\varepsilon_{IR})
+{1\over\textbf{T}_j^2}\Gamma_j(m_j,\varepsilon_{IR})
+{1\over\textbf{T}_a^2}\frac{\gamma_a}{\varepsilon_{IR}}\bigg]
\nonumber\\&& +\textbf{T}_a\cdot\textbf{T}_b\bigg[\left(
\frac{\mu_r^2}{s_{ab}}\right)^{\varepsilon}\left(\frac{1}
{\varepsilon_{IR}^2}+{1\over\textbf{T}_a^2}\frac{\gamma_a}
{\varepsilon_{IR}}\right)\bigg]+(a\leftrightarrow b)\bigg\},
\end{eqnarray}
with
\begin{eqnarray}
\mathcal{V}(s_{jk},m_j,m_k;\varepsilon_{IR})&=&{1\over
v_{jk}}\left(\frac{Q_{jk}^2}{s_{jk}}\right)^{\varepsilon}\times
\left( 1-{1\over 2}\rho_j^{-2\varepsilon}-{1\over
2}\rho_k^{-2\varepsilon}\right)
{1\over \varepsilon_{IR}^2},\nonumber\\
\Gamma_j(0,\varepsilon_{IR})&=&\frac{\gamma_j}{\varepsilon_{IR}},\quad
\Gamma_j(m_j\neq 0,\varepsilon_{IR})=\frac{C_F}{\varepsilon_{IR}},
\end{eqnarray}
where $C_F=4/3$, $\gamma_q=2$, and $\gamma_g=11/2-n_f/3$. And
$s_{jk}$, $Q_{jk}^2$, $v_{jk}$, and $\rho_n$ are kinematic variables
defined as follows
\begin{eqnarray}
s_{jk}&=&2p_jp_k,\quad Q_{jk}^2=s_{jk}+m_j^2+m_k^2,\quad
v_{jk}=\sqrt{1-\frac{m_j^2m_k^2}{(p_jp_k)^2}}, \nonumber\\
\rho_n&=&\sqrt{\frac{1-v_{jk}+2m_n^2/(Q_{jk}^2-m_j^2-m_k^2)}
{1+v_{jk}+2m_n^2/(Q_{jk}^2-m_j^2-m_k^2)}}\quad (n=j,k).
\end{eqnarray}
When inserting Eq.~(\ref{eq3}) into the LO amplitudes for the $q\bar
q$ and $qg$ subprocesses as shown in Eq.~(\ref{eq4}), we can see
that the IR divergences, including the $1/\epsilon_{\rm IR}$ terms,
can be written as combinations of the LO color correlated squared
amplitudes and all the IR divergences from the virtual corrections
in Eq.~(\ref{vir}) are canceled exactly, as we expected.

\section{Numerical Results}\label{s4}
In the Lagrangian $\mathcal{L}_{NP}$, there are totally nine free
parameter $C_{\rm LL}^1$, $C_{\rm LR}^1$, $C_{\rm RL}^1$, $C_{\rm RR}^1$,
$C_{\rm LL}^8$, $C_{\rm LR}^8$, $C_{\rm RL}^8$, $C_{\rm RR}^8$ and $\Lambda$. If we
include left-handed top quark $t_L$ in the $\mathcal{L}_{NP}$, it is
suitable to work in $SU(2)_L$ doublet of the third-generation quarks
$(t_{\rm L},b_{\rm L})^T$. However, the Wilson coefficients $C_{\rm LL}^1$, $C_{\rm
LL}^8$, $C_{\rm RL}^1$ and $C_{\rm RL}^8$ are highly constrained by
the LEP data for the ratio of $b\bar b$ to hadron
production~\cite{Nakamura:2010zzi}:
\begin{eqnarray}
  B_b=0.121629\pm0.00066,
\end{eqnarray}
which agrees well with the SM prediction. Thus, for simplicity we
choose $C_{\rm LL}^1=C_{\rm LL}^8=C_{\rm RL}^1=C_{\rm
RL}^8=0$~\cite{Jung:2009pi}. Up to $\mathcal{O}(\as^2/\Lambda^2)$,
contributions from color singlet operators due to mixing effects are
much less than contributions from color octet operators, so we only
consider color octet interactions. As a result, in the numerical
calculations there are only three free NP parameters in the
Lagrangian, i.e. $C_{\rm LR}^8$, $C_{\rm RR}^8$
and $\Lambda$.\\
\indent Top quark mass is taken to be $m_t=172.5~{\rm GeV}$.
 We choose CTEQ6L and CTEQ6M
PDF sets~\cite{Pumplin:2002vw} and the associated $\as$ functions
for LO and NLO calculation, respectively. Both the renormalization
and factorization scales are fixed to the top quark mass unless
specified. We have used the modified
MadDipole~\cite{Frederix:2008hu} package for the real corrections.

\begin{figure}[H]
  \subfigure{
    \begin{minipage}[b]{0.5\textwidth}
      \begin{center}
     \scalebox{0.45}{\includegraphics*[0,0][550,500]{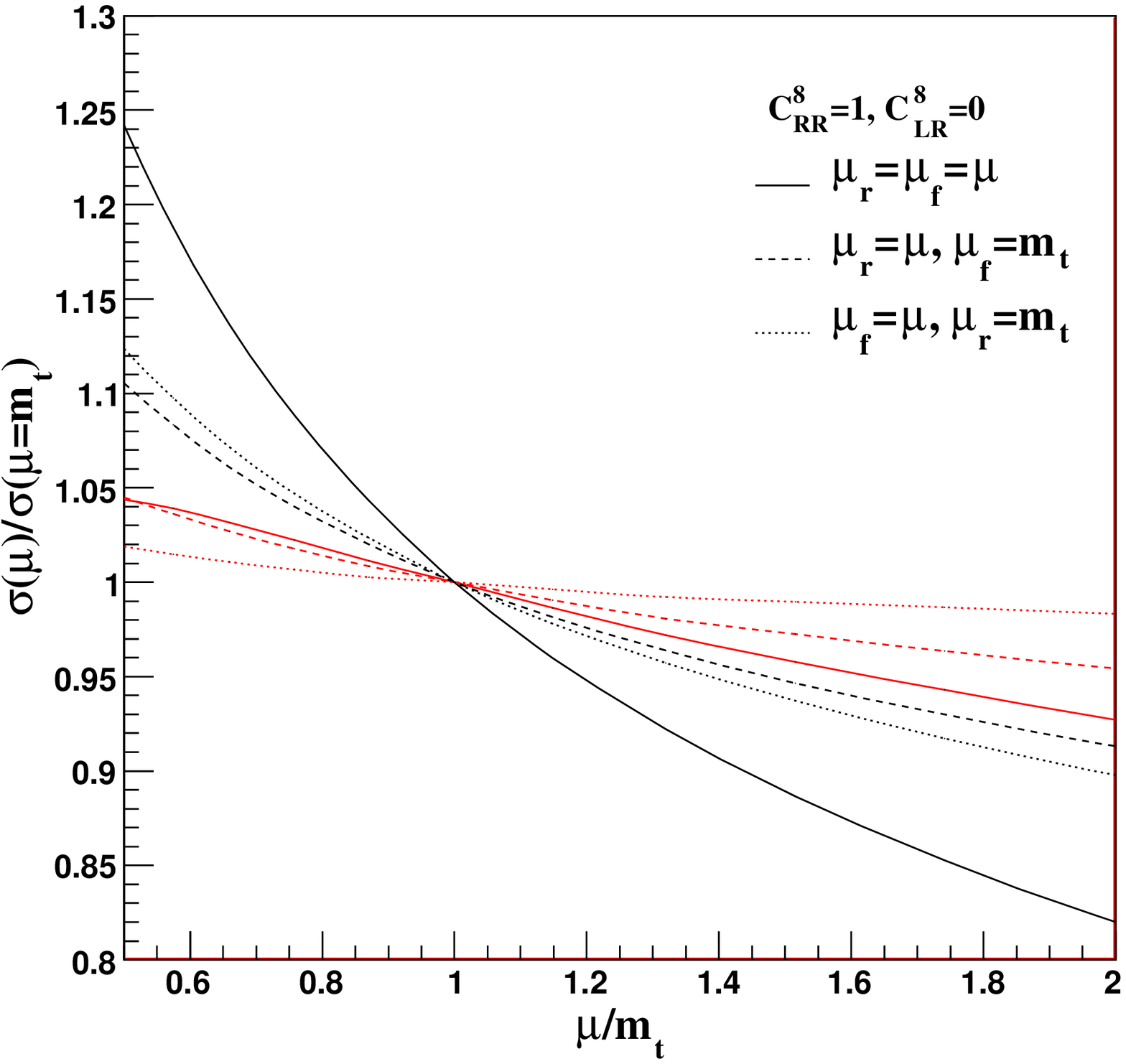}}
      \end{center}
    \end{minipage}}
  \subfigure{
    \begin{minipage}[b]{0.55\textwidth}
      \begin{center}
     \scalebox{0.45}{\includegraphics*[0,0][550,500]{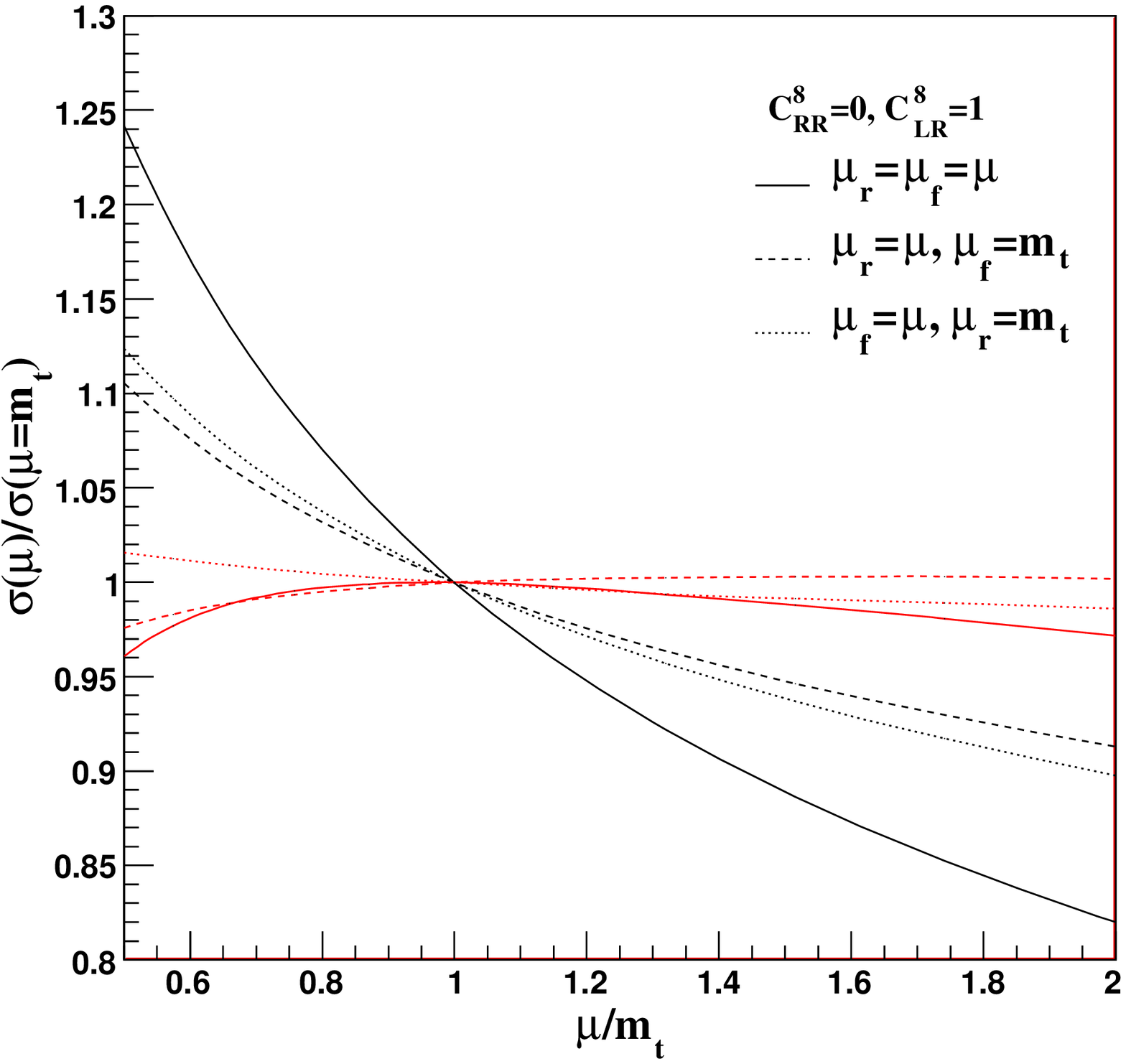}}
      \end{center}
    \end{minipage}}\vspace{-2mm}
    \caption{\label{fig:scale_dep} Scale dependence of the total cross
section at the Tevetron, the black and the red lines represent LO
and NLO results, respectively.}
\end{figure}

\subsection{Scale dependence}
In Fig.~\ref{fig:scale_dep} we show the scale dependence of the LO
and NLO total cross section at the Tevatron for three cases: (1) the
renormalization scale dependence $\mu_r=\mu,\ \mu_f=m_t$, (2) the
factorization scale dependence $\mu_r=m_t,\ \mu_f=\mu$, and (3)
total scale dependence $\mu_r=\mu_f=\mu$. It can be seen that the
NLO corrections reduce the scale dependence significantly for all
three cases, which makes the theoretical predictions more reliable.

\subsection{Tevatron constraints}

\begin{figure}[h!]
\begin{centering}
\includegraphics[scale=0.4]{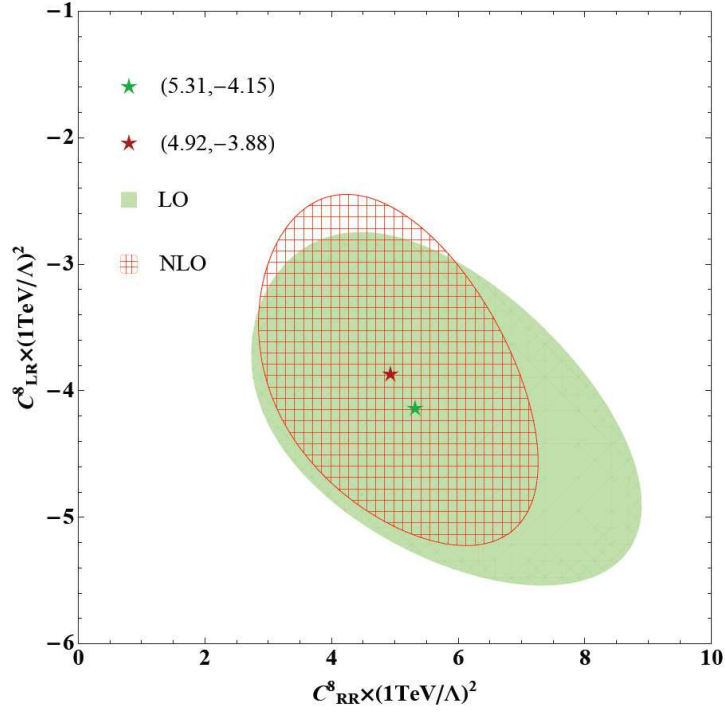}
\par\end{centering}
\caption{\label{fig:octet_para_scale} Values of $C_{RR}^8$ and
$C_{LR}^8$ allowed by the Tavetron data at 1$\sigma$ CL:
$\sigma_{t\bar t}$=(7.50 $\pm$ 0.48)pb and
$A_{\textrm{FB}}\left(m_{t\bar t} > 450 ~
\textrm{GeV}\right)$=0.475$\pm$0.114. The green star $\rm
(5.31,-4.15)$ and red star $\rm (4.92,-3.88)$ represent the BFPs at
LO and NLO level, respectively.}
\end{figure}

$A_{\textrm{FB}}$ of top quark pair productions is defined as
\begin{eqnarray}
  A_{\textrm{FB}}&=&\frac{\sigma_\textrm{F} - \sigma_\textrm{B}}{\sigma_\textrm{F} +
  \sigma_\textrm{B}}\nno\\
  &=&A_{\textrm{FB}}^{\textrm{NP}} \times R +
  A_{\textrm{FB}}^{\textrm{SM}}\times( 1 - R )\nno
\end{eqnarray}
where
\begin{eqnarray}
  A_{\textrm{FB}}^{\textrm{NP}}&=&(\sigma_\textrm{F}^{\textrm{NP}} -
  \sigma_\textrm{B}^{\textrm{NP}})/(\sigma_\textrm{F}^{\textrm{NP}} +
  \sigma_\textrm{B}^{\textrm{NP}}),\nno\\
  A_{\textrm{FB}}^{\textrm{SM}}&=&(\sigma_\textrm{F}^{\textrm{SM}} -
  \sigma_\textrm{B}^{\textrm{SM}})/(\sigma_\textrm{F}^{\textrm{SM}} +
  \sigma_\textrm{B}^{\textrm{SM}}),\nno\\
  R&=&\sigma_{\textrm{tot}}^{\textrm{NP}}/(\sigma_{\textrm{tot}}^{\textrm{SM}} + \sigma_{\textrm{tot}}^{\textrm{NP}})
\end{eqnarray}
are the asymmetries induced by NP and SM, and $R$ is the fraction of
NP contribution to the total cross section. $\sigma_{\textrm{F}}$
and $\sigma_{\rm B} $ denote the total cross sections in the
forward(F) and backward(B) rapidity regions, respectively. Up to
order $O(\as^2/\Lambda^2)$, total cross sections induced by NP can
be written as
\begin{eqnarray}\label{paracslo}
  \sigma^{\textrm{NP}}_{\textrm{LO}} &=& \left[
  (0.428^{+0.103}_{-0.076})C_{\rm RR}^8 + (0.428^{+0.101}_{-0.075})C_{\rm LR}^8\right]\left(\frac{1\text{TeV}}
  {\Lambda}\right)^2~\textrm{pb},
\end{eqnarray}
\begin{eqnarray}\label{paracsnlo}
  \sigma^{\textrm{NP}}_{\textrm{NLO}} &=& \left[
  (0.442^{+0.018}_{-0.032})C_{\rm RR}^8 + (0.435^{+0.022}_{-0.022})C_{\rm LR}^8\right]\left(\frac{1\text{TeV}}
  {\Lambda}\right)^2~\textrm{pb},
\end{eqnarray}
and the difference of the cross section in the forward and backward
rapidity regions can be written as
\begin{eqnarray}\label{paraafblo}
  \left[ \sigma_{\textrm{F}}^{\textrm{NP}} -
  \sigma_{\textrm{B}}^{\textrm{NP}}\right]^{m_{t\bar t} > 450 ~ \textrm{GeV}}_{\textrm{LO}} &=&
  \left[ (0.118^{+0.031}_{-0.023})C_{\rm RR}^8-(0.118^{+0.031}_{-0.023})C_{\rm LR}^8 \right]
  \left(\frac{1\text{TeV}}{\Lambda}\right)^2~\textrm{pb},
\end{eqnarray}
\begin{eqnarray}\label{paraafbnlo}
  \left[ \sigma_{\textrm{F}}^{\textrm{NP}} -
  \sigma_{\textrm{B}}^{\textrm{NP}} \right]^{m_{t\bar t} > 450 ~ \textrm{GeV}}_{\textrm{NLO}} &=&
  \left[ (0.149^{+0.003}_{-0.009})C_{\rm RR}^8-(0.103^{+0.008}_{-0.025})C_{\rm LR}^8 \right]
\left(\frac{1\text{TeV}}{\Lambda}\right)^2~\textrm{pb},
\end{eqnarray}
where the errors are obtained by varying the scale between $\mu_r =
\mu_f = m_t/2$ and $\mu_r = \mu_f = 2m_t$. From the expressions
Eqs.(\ref{paracslo} -- \ref{paraafbnlo}) we can see that NLO
corrections reduce the dependence of
$\sigma_{\textrm{F}}^{\textrm{NP}} -
\sigma_{\textrm{B}}^{\textrm{NP}}$ and $\sigma^{\textrm{NP}}$ on the
renormalization and factorization scales significantly.\\
\indent In Fig.~\ref{fig:octet_para_scale}, we show the allowed
region in the $(C_{\rm RR}^8 ,C_{\rm LR}^8)$ plane that is
consistent with the Tevatron data~\cite{Aaltonen:2011kc}:
\begin{eqnarray}
\sigma_{t\bar{t}}^{\rm EX}&=&(7.50 \pm 0.48)\textrm{pb}, \nno\\
\rm{A_{FB}^{EX}}&=&0.475 \pm 0.114,~~~~{\rm for}~m_{t\bar t} > 450~
\rm{GeV}.
\end{eqnarray}
We use Monte Carlo programm MCFM~\cite{Campbell:1999ah} to get the
cross section of the gluon fusion channel $gg\to t\bar t$ at the NLO
QCD level. As for the process of $q \bar q\to t\bar t$, we have
checked our value with the results given by MCFM at QCD NLO level,
which are well consistent in the range of Monte Carlo integration error.
Combining the contributions of these two channels
we have the total cross section of $t\bar t$ production
\begin{eqnarray}
\sigma_{t\bar{t}}^{\textrm{SM}}&=&7.00^{+0.36}_{-0.76}~\textrm{pb},
\end{eqnarray}
where we have considered scale uncertainty in the calculations. For
consistency we have used the SM predicted values of $A_{\rm
FB}\left( m_{t\bar t} \geq 450 \rm{GeV} \right)=0.088$ at NLO QCD
level, although next-to-next-to-leading logarithmic (NNLL) SM QCD
results are available~\cite{Ahrens:2011uf}. In
Fig.~\ref{fig:octet_para_scale}, green and red regions correspond to
NP LO and NLO results at 1$\sigma$ confidence level(CL), where we
have considered theoretical and experimental uncertainty in the
total cross section and only consider experimental uncertainty in
the $A_{\rm FB}$ calculation. It can be seen that NLO corrections
obviously change the allowed region of $C_{\textrm{RR}}^8$ and
$C_{\textrm{RR}}^1$. The green star $\rm (5.31,-4.15)$ and red star
$\rm (4.92,-3.88)$ represent the best-fit point(BFP) at LO and NLO
level, respectively, from which we can see higher order corrections
reduce the BFP by about 7\%. The $t\bar t$ total cross sections
induced by NP at the NLO BFP $\rm (4.92,-3.88)$ are
\begin{eqnarray}
  \sigma^{\rm{NP}}_{t\bar t,\rm{LO}}&=&0.445~\rm{pb},\nno\\
  \sigma^{\rm{NP}}_{t\bar t,\rm{NLO}}&=&0.497~\rm{pb},
\end{eqnarray}
where the K factor is about 1.12. $A_{\rm FB}$ containing NP
contributions at the NLO BFP are shown together in
Table~\ref{tab:bestfit}. The NLO QCD corrections to $A_{\rm FB}$ can
reach about 10\%, and the theoretical predictions containing
NP NLO effects are consistent with experimental results at 2$\sigma$ CL.\\
\begin{table*}[h!]
\begin{center}
\scalebox{1}[0.9] {\begin{tabular}{c|c|c}
  \hline
  \hline
  &SM NLO QCD + NP LO &SM NLO QCD + NP NLO  \\
  \hline
  $A_{\rm FB}^{p\bar p}$  & 0.175  & 0.189 ($\sim$ 0.7 $\sigma$) \\
  $A_{\rm FB}^{t\bar t}$  & 0.252  & 0.275 ($\sim$ 1.6 $\sigma$) \\
  $A_{\rm FB}^{t\bar t}\left(m_{t\bar t} < 450~{\rm GeV}\right)$  & 0.132  & 0.136 ($\sim$ 1.6 $\sigma$) \\
  $A_{\rm FB}^{t\bar t}\left(m_{t\bar t} > 450~{\rm GeV}\right)$  & 0.452& 0.475 ($\sim$ 0 $\sigma$) \\
  $A_{\rm FB}^{t\bar t}\left(|\Delta y|<1\right)$  & 0.170 & 0.161  ($\sim$ 0.9 $\sigma$) \\
  $A_{\rm FB}^{t\bar t}\left(|\Delta y|>1\right)$  & 0.719 &  0.681 ($\sim$ 0.3 $\sigma$) \\
  \hline
  \hline
\end{tabular}}
\end{center}
\caption{\label{tab:bestfit} $A_{\rm FB}$ with $\rm
C_{RR}^8(1TeV/\Lambda)^2 = 4.92 $ and $\rm C_{LR}^8(1TeV/\Lambda)^2
= -3.88 $ at the Tevtron, where $A_{\rm FB}^{p\bar p}$ and $A_{\rm
FB}^{t\bar t}$ are the $A_{\rm FB}$ in the lab frame and the $t\bar
t$ rest frame, respectively, and $\Delta y = y_{t} - y_{\bar t}$ is
the difference of rapidities of the top and antitop quarks. Here we
list the CL when containing NP effects at NLO level.}
\end{table*}
\begin{figure}[h!]
\begin{centering}
\includegraphics[scale=0.6]{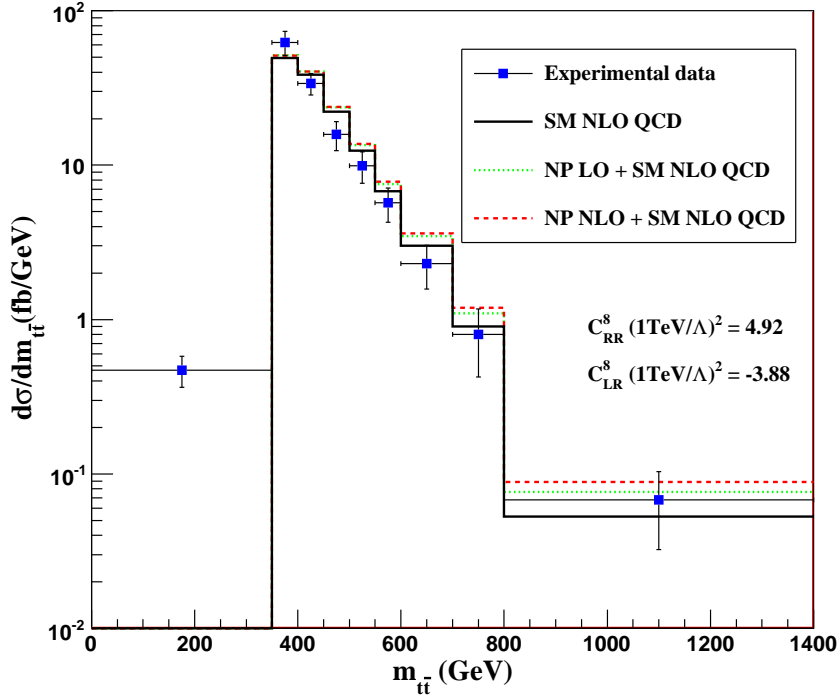}
\par\end{centering}
\caption{\label{fig:mtt_tev} Differential cross sections $d\sigma/d
m_{t\bar{t}}$ as a function of  $m_{t\bar t}$ at the NLO BFP $\rm
(4.92,-3.88)$. Here "Experimental data" is $d\sigma/d m_{t\bar{t}}$
measured with 2.7 $fb^{-1}$ of integrated luminosity at the
Tevatron~\cite{Aaltonen:2009iz}. "SM NLO QCD" represents the results
in the SM QCD at NLO level. "NP LO + SM NLO QCD" and "NP NLO + SM
NLO QCD" stand for the predictions including NP effects up to
$\mathcal{O}(\as/\Lambda^2)$ and $\mathcal{O}(\as^2/\Lambda^2)$,
respectively.  }
\end{figure}
\indent In Fig.~\ref{fig:mtt_tev}, we show differential cross
section $d\sigma/d m_{t\bar{t}}$ when we consider NP effects at the
NLO BFP, from which we can see that higher
order corrections do not change the distribution very much.\\

\subsection{LHC predictions}

\begin{figure}[h!]
\begin{centering}
\includegraphics[scale=0.6]{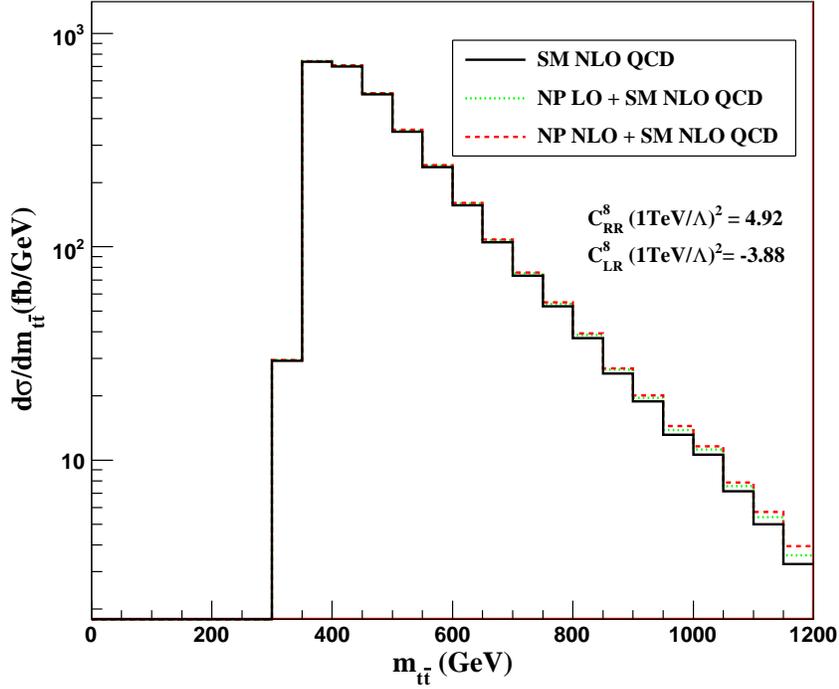}
\par\end{centering}
\caption{\label{fig:mtt_lhc} Differential cross sections $d\sigma/d
m_{t\bar{t}}$  as a function of  $m_{t\bar t}$ at the LHC with
$\sqrt{S}=7~{\rm TeV}$.}
\end{figure}

\indent The process of top quark pair production has been measured
at the LHC, and the cross
section~\cite{Cristinziani:2011cu,Tosi:2011wd} is
\begin{eqnarray}
  \sigma_{t\bar t}^{\rm ATLAS}({\rm \sqrt{S}=7~TeV}) = 180 \pm 18 {\rm pb}, \nno\\
  \sigma_{t\bar t}^{\rm CMS}({\rm \sqrt{S}=7~TeV}) = 158 \pm 19 {\rm
  pb},
\end{eqnarray}
which is consistent with the SM predictions. The NP contributions at
the NLO BFP $\rm (4.92,-3.88)$ is about 3 pb, which is much smaller
than the experimental uncertainty. Thus, it is difficult to measure
the NP effects only through the cross section measurements. In
Fig.~\ref{fig:mtt_lhc}, we show differential cross section
$d\sigma/d m_{t\bar{t}}$ at the LHC when we consider NP effects at
the NLO BFP $\rm (4.92,-3.88)$, from which we can see that NP
contributions almost do not change the distribution.
\\
\indent Since the LHC is a proton-proton collider, which is
forward-backward symmetric, the $A_{\rm FB}$ defined at Tevatron can
not be directly applied to the proton-proton collider experiments at
the LHC. The $A_{\rm C}$ used by CMS~\cite{CMS-PAS-TOP-10-014} can
be written as
\begin{eqnarray}
  A_{\rm C}=\frac{\sigma(|\eta_{t}|-|\eta_{\bar t}|>0) - \sigma(|\eta_{t}|-|\eta_{\bar t}|<0)}
  {\sigma(|\eta_{t}|-|\eta_{\bar t}|>0) + \sigma(|\eta_{t}|-|\eta_{\bar
  t}|<0)},
\end{eqnarray}
where $\eta_{t}$ and $\eta_{\bar t}$ are pseudo rapidities of top
and antitop quark, respectively. Its value is measured to be
\begin{eqnarray}
  A_{\rm C}=-0.016 \pm 0.030({\rm stat.})^{+ 0.010}_{-0.019}({\rm syst.})
\end{eqnarray}
which is consistent with the SM predictions: $A_{\rm
C}=0.013(11)$~\cite{CMS-PAS-TOP-10-014,Ferrario:2009ee,Ferrario:2008wm}.
The $A_{\rm C}$ induced by NP interactions at the NP NLO BFP $\rm
(4.92,-3.88)$ is 0.063, which is about 5 times larger
than SM predictions, and also consistent with the CMS data at about 2$\sigma$ CL.\\
\begin{figure}[h!]
\begin{center}
\includegraphics[scale=0.3]{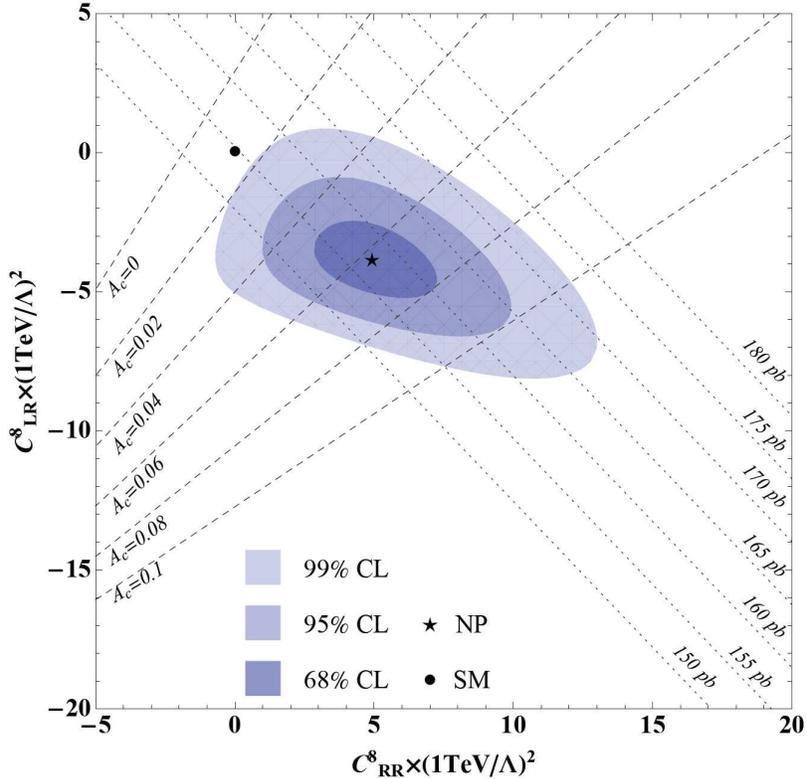}
\end{center} \vspace{-7.5mm}
\begin{center}
\parbox{15.5cm}{\caption{\label{fig:para_tota} Results of a combined fit to
$\sigma_{t\bar t}$ and the value of $A_{\rm FB}$ allowing for NP at
different CLs. The blue contours from dark to light indicate the
experimentally favored region of 68\%, 95\% and 99\% CL in the
($C_{RR}^8,C_{LR}^8$) plane. The black doted and dashed lines
respectively represent the value of $t\bar t$ cross section and the
$A_{\rm C}$ at the LHC with $\sqrt{S}=7$ TeV. The black dot and
black star represent the SM point and the NP NLO BFP.}}
\end{center}
\end{figure}
\indent In Fig.~\ref{fig:para_tota}, we show the results of a
combined fit to the $t\bar t$ data in the presence of NP at
different CLs. The blue contours from dark to light indicate the
experimentally preferred region of 68\%, 95\% and 99\% CL in the
($C_{RR}^8,C_{LR}^8$) plane. The black dot represent the SM point
$(0,0)$, and the black star represent the NP NLO BFP $\rm
(4.92,-3.88)$. The black doted and dashed lines respectively
represent the value of $t\bar t$ cross section and the $A_{\rm C}$
at the LHC with $\sqrt{S}=7$ TeV. From the location of the blue
area, one finds that $A_{\rm C}$ predicted by NP is obviously much
larger than SM predictions, and LHC has potential to find this
difference when the measurement precision increases.

\section{Conclusions}\label{s5}
In conclusion, we have investigated $A_{\rm FB}$ and total cross
section of top quark pair production induced by dimension-six four
quark operators at the Tevatron up to
$\mathcal{O}(\as^2/\Lambda^2)$. Our results show that, NLO QCD
corrections can change $A_{\rm FB}$ and the total cross section by
about 10\%. Moreover, NLO QCD corrections reduce the dependence of
$A_{\rm FB}$ and the total cross sections  on the renormalization
and factorization scales significantly, which lead to increased
confidence on the theoretical predictions. We also evaluate total
cross section and $A_{\rm C}$ induced by these operators at the LHC
up to $\mathcal{O}(\as^2/\Lambda^2)$, for the parameter space
allowed by the Tevatron data. We find that the value of $A_{\rm C}$
induced by these operators is much larger than SM predictions, and
LHC has potential to discover these NP effects when the measurement
precision increases.

\begin{acknowledgments}
We would like to thank Rikkert Frederix, Kouhei Hasegawa and
Sven-Olaf Moch for helpful discussion.  This work was supported in
part by the National Natural Science Foundation of China, under
Grants No. 11021092 and No. 10975004.
\end{acknowledgments}

\appendix
\section{FINITE TERMS IN VIRTUAL CORRECTIONS}\label{a1}
In this appendix, we collect explicit expressions of finite terms in
virtual corrections. Analytical continuation for the Mandelstam
variables are defined as
\begin{eqnarray}
  s &\to& s+i\varepsilon,\nno\\
  u &\to& u+i\varepsilon,\nno\\
  t &\to& t+i\varepsilon.
\end{eqnarray}
For simplicity, we introduce the following abbreviations.
\begin{eqnarray}
c_t=\frac{m_t^2-t}{t},~~c_u=\frac{m_t^2-u}{u},
\end{eqnarray}
\begin{eqnarray}
  y_1(t) &=& \ln\frac{t_1}{\mu_r^2} + \frac{t_1}{2t} - 1 ,\nno\\
  y_2(t) &=& \ln\frac{m_t^2}{\mu_r^2} +  \frac{t_1}{t} - 1,\nno\\
  y_3(t) &=& \ln\frac{t_1}{\mu_r^2} - \left( \frac{t_1}{t} + 1\right),\nno\\
  y_4(t) &=& \ln\frac{m_t^2}{\mu_r^2} - 2\left( \frac{t_1}{t} + 2\right),
\end{eqnarray}
\begin{eqnarray}
f_1(t)&=&\frac{3}{8t}\left(\frac{2}{c^2_t}+\frac{2}{c_t}+\ln
\frac{t_1}{m_t^2}\right),\nno\\
f_2(t)&=&\frac{3}{4}\left(1+\frac{1}{c_t}\right)\left(1-c_t\ln
\frac{t_1}{m_t{}^2}\right),\nno\\
f_3(t)&=&-\frac{1}{2}y_1{}^2(t)+\frac{1}{4}y_2{}^2(t)-\frac{1}{8}c^2_t-\frac{1}{2c_t}-\frac{\pi
^2}{12}-\frac{5}{4}+\text{Li}_2\left(\frac{-t}{t_1}\right),\nno\\
f_4(t)&=&-\frac{1}{2}y_3{}^2(t)+\frac{1}{4}y_4{}^2(t)-\frac{1}{2}c^2_t-3c_t-\frac{1}{c_t}-\frac{\pi
^2}{12}-\frac{3}{2}+\text{Li}_2\left(\frac{-t}{t_1}\right),\nno\\
f_5(t,u)&=&\frac{3}{2}f_3(t)-\frac{3}{2}f_4(u),
\end{eqnarray}
\begin{eqnarray}
h_1(s)&=&\frac{1}{32}\left(\beta +\frac{1}{\beta }\right)\ln
\left(\frac{\beta +1}{\beta -1}\right)\left(2\text{ln$\beta $}-3+\ln
\frac{m_t^2}{\mu_r^2}+\ln \left(\frac{-s}{\mu_r
^2}\right)\right),\nno\\
h_2(s)&=&\frac{1}{16}\left(\beta +\frac{1}{\beta
}\right)\left(\text{Li}_2\left(\frac{\beta -1}{2\beta
}\right)-\text{Li}_2\left(\frac{\beta +1}{2\beta }\right)\right),\nno\\
h_3(s)&=&-\frac{\beta ^3}{16}\ln \left(\frac{\beta +1}{\beta
-1}\right)+\frac{\beta ^2}{8}+\frac{3\beta }{16}\ln
\left(\frac{\beta +1}{\beta -1}\right)+\frac{1}{16}\ln
^2\left(\frac{-s}{\mu_r^2}\right),\nno\\
&&-\frac{21}{16}\ln \frac{m_t^2}{\mu_r ^2}-\frac{3-2n_f}{16}\ln
\left(\frac{-s}{\mu_r
^2}\right)-\frac{5n_f-61}{24},\nno\\
h_4(s)&=&-\frac{\beta ^3}{16}\ln \left(\frac{\beta +1}{\beta
-1}\right)+\frac{\beta ^2}{8}+\frac{5\beta }{32}\ln
\left(\frac{\beta +1}{\beta -1}\right)+\frac{1}{8}\ln
\frac{m_t^2}{\mu_r^2},\nno\\
&&-\frac{1}{3}+\frac{1}{32\beta }\ln
\left(\frac{\beta +1}{\beta -1}\right),\nno\\
h_5(s)&=&\frac{1}{16s \beta }\ln \left(\frac{\beta +1}{\beta
-1}\right),\nno\\
h_6(s)&=&n_f\left(\frac{1}{8}\ln \left(\frac{-s}{\mu_r
^2}\right)-\frac{5}{24}\right),\nno\\
h_7(s)&=&-\frac{1}{2}\ln ^2\left(\frac{-s}{\mu_r
^2}\right)+\frac{3}{2}\ln \left(\frac{-s}{\mu_r^2}\right)-2\ln
\frac{m_t^2}{\mu_r^2}-\frac{1}{2},\nno\\
h_8(s)&=&\frac{1}{4}\left(\beta -\frac{1}{\beta }\right)\ln
\left(\frac{\beta +1}{\beta -1}\right),\nno\\
h_9(s)&=&-8h_1(s)-8h_2(s)+h_7(s),
\end{eqnarray}
\begin{eqnarray}
g_1\left(s,f_3(t),f_4(u)\right)&=&\frac{7}{4}f_3(t)+\frac{1}{2}f_4(u)+h_1(s)+h_2(s)+h_3(s),\nno\\
g_2(s,t)&=&\frac{7}{6}f_2(t)+h_4(s),\nno\\
g_3(s,u)&=&\frac{1}{3}f_2(u)+h_4(s),\nno\\
g_4(s,t)&=&-\frac{7}{6}f_1(t)-h_5(s),\nno\\
g_5(s,u)&=&-\frac{1}{3}f_1(u)-h_5(s).
\end{eqnarray}

First we list the amplitude for helicity $(+ - + +)$,
\begin{eqnarray}
  i\mathcal{M}_{\rm virt}^{\rm NP,fin}(+ - + +) = \frac{i\as m}{\pi\Lambda^2}
  C_F \big[ \mathcal{M}_{\rm ss}(+ - + +) \mathcal{C}_1 + \mathcal{M}_{\rm so}(+ - + +) \mathcal{C}_8
  \nno\\
  + \mathcal{M}_{\rm os}(+ - + +) \mathcal{C}_1 + \mathcal{M}_{\rm oo}(+ - + +) \mathcal{C}_8
  \big],
\end{eqnarray}
where
\begin{eqnarray}
\mathcal{M}_{\text{ss}}(+-\text{++}) &=& C^1_{\text{RL}}
\left[h_8(s)\mathcal{M}_1
+ h_9(s)\mathcal{M}_2 + 8h_5(s)\left(\mathcal{M}_1{}^{(m)} + \mathcal{M}_2{}^{(m)}\right)\right]\nno\\
&& + C^1_{\text{RR}}\left[h_9(s)\mathcal{M}_1 +
h_8(s)\mathcal{M}_2+8h_5(s)\left(\mathcal{M}_1{}^{(m)} + \mathcal{M}_2{}^{(m)}\right)\right],\\
\mathcal{M}_{\text{so}}(+-\text{++}) &=& -C^1_{\text{RL}} \left[
f_2(u) \mathcal{M}_1 +
f_5(u,t)\mathcal{M}_2 - f_1(u)\left(\mathcal{M}_1{}^{(m)} + \mathcal{M}_2{}^{(m)}\right)\right]\nno\\
&&+C^1_{\text{RR}}\left[f_5(t,u)\mathcal{M}_1 + f_2(t)\mathcal{M}_2
- f_1(t)\left(\mathcal{M}_1{}^{(m)}
+ \mathcal{M}_2{}^{(m)}\right)\right],\\
\mathcal{M}_{\text{os}}(+-\text{++}) &=& -\frac{2}{9}C^8_{\text{RL}}
\left[ f_2(u) \mathcal{M}_1 +
f_5(u,t)\mathcal{M}_2 - f_1(u)\left(\mathcal{M}_1{}^{(m)} + \mathcal{M}_2{}^{(m)}\right)\right]\nno\\
&&+\frac{2}{9}C^8_{\text{RR}}\left[f_5(t)\mathcal{M}_1 +
f_2(t)\mathcal{M}_2 - f_1(t)\left(\mathcal{M}_1{}^{(m)}
+ \mathcal{M}_2{}^{(m)}\right)\right],\\
\mathcal{M}_{\text{oo}}(+-\text{++}) &=&
C^8_{\text{RL}}\left[g_3(s,u)\mathcal{M}_1+g_1\left(s,f_4(t),f_3(u)\right)\mathcal{M}_2
+g_5(s,u)\left(\mathcal{M}_1{}^{(m)} + \mathcal{M}_2{}^{(m)}\right)\right]\nno\\
&&+C^8_{\text{RR}}\left[g_1\left(s,f_3(t),f_4(u)\right)\mathcal{M}_1+g_2(s,t)\mathcal{M}_2
+g_4(s,t)\left(\mathcal{M}_1{}^{(m)}+\mathcal{M}_2{}^{(m)}\right)\right]\nno\\
&&+C^8_{\text{LR}}h_6(s)\mathcal{M}_1+C^8_{\text{LL}}h_6(s)\mathcal{M}_2.
\end{eqnarray}
Similarly, amplitude for helicity $(- + + +)$ can be written as
\begin{eqnarray}
  i\mathcal{M}_{\rm virt}^{\rm NP,fin}(- + + +) = \frac{i\as m}{\pi\Lambda^2}
  C_F \big[ \mathcal{M}_{\rm ss}(- + + +) \mathcal{C}_1 + \mathcal{M}_{\rm so}(- + + +) \mathcal{C}_8
  \nno\\
  + \mathcal{M}_{\rm os}(- + + +) \mathcal{C}_1 + \mathcal{M}_{\rm oo}(- + + +) \mathcal{C}_8
  \big],
\end{eqnarray}
where
\begin{eqnarray} \mathcal{M}_{\text{ss}}(-\text{++}+) &=&
C^1_{\text{LR}} \left[h_9(s)\mathcal{M}_3 +
h_8(s)\mathcal{M}_4 - 8h_5(s)\left(\mathcal{M}_3{}^{(m)} + \mathcal{M}_4{}^{(m)}\right)\right]\nno\\
&&+C^1_{\text{LL}} \left[h_8(s)\mathcal{M}_3 + h_9(s)M_4 -
8h_5(s)\left(\mathcal{M}_3{}^{(m)} +
\mathcal{M}_4{}^{(m)}\right)\right],\\
\mathcal{M}_{\text{so}}(-\text{++}+) &=& -C^1_{\text{LR}} \left[
f_5(u,t)\mathcal{M}_3 + f_2(u)\mathcal{M}_4 + f_1(u)
\left(\mathcal{M}_3{}^{(m)} + \mathcal{M}_4{}^{(m)}\right)\right]\nno\\
&&+C^1_{\text{LL}}\left[f_2(t)\mathcal{M}_3 +
f_5(t,u)\mathcal{M}_4+f_1(t)\left(\mathcal{M}_3{}^{(m)} +
\mathcal{M}_4{}^{(m)}\right)\right],\\
\mathcal{M}_{\text{os}}(-\text{++}+) &=& -\frac{2}{9}C^8_{\text{LR}}
\left[ f_5(u,t)\mathcal{M}_3 + f_2(u)\mathcal{M}_4 + f_1(u)
\left(\mathcal{M}_3{}^{(m)} + \mathcal{M}_4{}^{(m)}\right)\right]\nno\\
&&+\frac{2}{9}C^8_{\text{LL}}\left[f_2(t)\mathcal{M}_3 +
f_5(t,u)\mathcal{M}_4+f_1(t)\left(\mathcal{M}_3{}^{(m)} +
\mathcal{M}_4{}^{(m)}\right)\right],\\
\mathcal{M}_{\text{oo}}(-\text{++}+) &=&
C^8_{\text{LL}}\left[g_2(s,t)\mathcal{M}_3 +
g_1\left(s,f_3(t),f_4(u)\right)\mathcal{M}_4-
g_4(s,t)\left(\mathcal{M}_3{}^{(m)} + \mathcal{M}_4{}^{(m)}\right)\right]\nno\\
&& +
C^8_{\text{LR}}\left[g_1\left(s,f_4(t),f_3(u)\right)\mathcal{M}_3 +
g_3(s,u)\mathcal{M}_4 - g_5(s,u)\left(\mathcal{M}_3{}^{(m)} + \mathcal{M}_4{}^{(m)}\right)\right]\nno\\
&&+C^8_{\text{RL}}h_6(s)\mathcal{M}_4 +
C^8_{\text{RR}}h_6(s)\mathcal{M}_3.
\end{eqnarray}

\section{evolution equation of Wilson coefficients}\label{a2}
In this appendix we present the evaluation equations of the Wilson
coefficients $C^{1}_{\rm AB}$ and $C^{8}_{\rm AB}$, expanded to
$\mathcal{O}(\as)$.
\begin{eqnarray}
   C_{\rm LL}^1(\mu_r)&=& C_{\rm LL}^1(m_t)-0.0397887
  C_F\as\ln\frac{m_t}{\mu_r}  C_{\rm LL}^8(m_t),\\
   C_{\rm LL}^8(\mu_r)&=&{\rm{ C_{LL}^8}}(m_t)-0.0497359 C_F \as
  \ln\frac{m_t}{\mu_r} C_{\rm RL}^8(m_t)\nno\\
  &&- 0.179049 C_F \as \ln\frac{m_t}{\mu_r}  C_{\rm LL}^1(m_t)\nno\\
  &&-0.00994718 C_F \as \ln\frac{m_t}{\mu_r}  C_{\rm LR}^8(m_t),\\
   C_{\rm LR}^1(\mu_r)&=& C_{\rm LR}^1(m_t) + 0.0397887 C_F \as
  \ln\frac{m_t}{\mu_r}  C_{\rm LR}^8(m_t),\\
   C_{\rm LR}^8(\mu_r)&=& C_{\rm LR}^8(m_t) + 0.149208 C_F \as
  \ln\frac{m_t}{\mu_r} C_{\rm LR}^8(m_t)\nno\\
  &&- 0.0497359 C_F \as \ln\frac{m_t}{\mu_r}
   C_{\rm RR}^8(m_t)\nno\\
  &&-0.00994718 C_F \as \ln\frac{m_t}{\mu_r}
   C_{\rm LL}^8(m_t) \nno\\
  &&+0.179049 C_F \as \ln\frac{m_t}{\mu_r}
   C_{\rm LR}^1(m_t),\\
   C_{\rm RL}^1(\mu_r)&=& C_{\rm RL}^1(m_t)+0.0397887 C_F \as \ln\frac{m_t}{\mu_r}
   C_{\rm RL}^8(m_t),\\
   C_{\rm RL}^8(\mu_r)&=& C_{\rm RL}^8(m_t)+0.149208 C_F \as
  \ln\frac{m_t}{\mu} C_{\rm RL}^8(m_t) \nno\\
  &&- 0.0497359  C_F \as \ln\frac{m_t}{\mu_r}
   C_{\rm LL}^8(m_t)\nno\\
  &&-0.00994718 C_F \as \ln\frac{m_t}{\mu_r}
   C_{\rm RR}^8(m_t)\nno\\
  &&+0.179049 C_F \as \ln\frac{m_t}{\mu_r}
   C_{\rm RL}^1(m_t),\nno\\
   C_{\rm RR}^1(\mu_r)&=& C_{\rm RR}^1(m_t)-0.0397887  C_F \as \ln\frac{m_t}{\mu_r}
   C_{\rm RR}^8(m_t),\\
   C_{\rm RR}^8(\mu_r)&=& C_{\rm RR}^8(m_t) - 0.0497359 C_F \as
  \ln\frac{m_t}{\mu_r} C_{\rm LR}^8(m_t)\nno\\
  &&- 0.00994718 C_F \as \ln\frac{m_t}{\mu_r} C_{\rm RL}^8(m_t)\nno\\
  &&- 0.179049 C_F \as \ln\frac{m_t}{\mu_r} C_{\rm RR}^1(m_t).
\end{eqnarray}

\bibliography{FBApaper}

\end{document}